\documentclass[prl,twocolumn,amsmath,amssymb,superscriptaddress]{revtex4-1} % ,linenumbers

\usepackage{times}
\usepackage{graphicx}% Include figure files
\usepackage{bm}% bold math
\usepackage[dvipsnames]{xcolor}
\usepackage[normalem]{ulem} % enables strikeout (\sout) without changing \emph
\usepackage{textcomp}
\usepackage{lineno}
\usepackage[sort&compress]{natbib}
\usepackage{hyperref}
\usepackage{xspace} % correctly handles spaces after macros
\usepackage{siunitx}
\usepackage{amsmath}
\usepackage{array}
\usepackage{enumitem}
\newcolumntype{P}[1]{>{\centering\arraybackslash}p{#1}}

\newcommand{\nbse}{NbSe$_2$~}
\newcommand{\wte}{WTe$_2$~}

\newcommand{\Dedge}{\Delta_{\rm{WTe_2}}^{\rm(edge)}}
\newcommand{\Dbulk}{\Delta_{\rm{WTe_2}}^{\rm(ML)}}
\newcommand{\DbulkBL}{\Delta_{\rm{WTe_2}}^{\rm(BL)}}

\newcommand{\murm}{%
	\ifmmode
	\mathchoice
	{\hbox{\normalsize\textmu}}
	{\hbox{\normalsize\textmu}}
	{\hbox{\scriptsize\textmu}}
	{\hbox{\tiny\textmu}}%
	\else
	\textmu
	\fi
}

\begin{document} 
	
	%\title{Proximity-induced superconducting gap in the quantum spin Hall edge state \\ of monolayer WTe$_2$}
	%\title{Coexistence of quantum spin Hall edge state and proximity-induced superconducting gap in monolayer 1T'-WTe$_2$}
	
	%\author{Felix L\"upke} %\email{These authors contributed equally.}
	%\author{Dacen Waters} %\email{These authors contributed equally.} 
	%\author{Sergio C. de la Barrera}
	%\author{Michael Widom}
	%\affiliation{Department of Physics, Carnegie Mellon University, Pittsburgh, PA 15213}
	%\author{David G. Mandrus}
	%\affiliation{Materials Science and Technology Division, Oak Ridge National Laboratory, Oak Ridge, TN 37831, USA}
	%\affiliation{Department of Materials Science and Engineering, University of Tennessee, Knoxville, TN 37996, USA}
	%\affiliation{Department of Physics and Astronomy, University of Tennessee, Knoxville, TN 37996, USA}
	%\author{Jiaqiang Yan}
	%\affiliation{Materials Science and Technology Division, Oak Ridge National Laboratory, Oak Ridge, TN 37831, USA}
	%\author{Randall M. Feenstra}
	%\author{Benjamin M. Hunt} %\email{bmhunt@andrew.cmu.edu} 
	%\affiliation{Department of Physics, Carnegie Mellon University, Pittsburgh, PA 15213}
	
	%\maketitle

	%%%%%%%%%%%%%%%%% END OF PREAMBLE %%%%%%%%%%%%%%%%

	
	% \baselineskip24pt
	
	% Make the title - Felix version
	%%%%%%%%%%%%%%%%%%%%%%%%%%%%%%%%%%%%%%%%%%%%%%%%%%%%%%%%%%%%%%%%%%%%
	
	%\maketitle
	
	%\begin{center}
	%\onecolumngrid
	\title{Proximity-induced superconducting gap in the quantum spin Hall edge state of monolayer WTe$_2$\\}
	%\vspace{0.5cm}
	\author{
		Felix L\"upke,$^{1\ast}$ Dacen Waters,$^{1\ast}$ Sergio C. de la Barrera,$^{1}$ Michael Widom,$^{1}$\\ David G. Mandrus,$^{2,3,4}$ Jiaqiang Yan,$^{2}$ Randall M. Feenstra,$^{1}$ and Benjamin M. Hunt$^{1\dagger}$\\
		\vspace{0.25cm}
		\normalsize{$^{1}$\textit{ Department of Physics, Carnegie Mellon University, Pittsburgh, PA 15213, USA}}\\
		\normalsize{$^{2}$\textit{Materials Science and Technology Division, Oak Ridge National Laboratory, Oak Ridge, TN 37831, USA}}\\
		\normalsize{$^{3}$\textit{Department of Materials Science and Engineering, University of Tennessee, Knoxville, TN 37996, USA}}\\
		\normalsize{$^{4}$\textit{Department of Physics and Astronomy, University of Tennessee, Knoxville, TN 37996, USA}}\\
		\normalsize{$^\ast$ These authors contributed equally}\\
		\normalsize{$^\dagger$E-mail: bmhunt@andrew.cmu.edu}
	}
	%\vspace{0.5cm}
	
	%\end{center}
	%\twocolumngrid
	
	\maketitle
	%%%%%%%%%%%%%%%%%%%%%%%%%%%%%%%%%%%%%%%%%%%%%%%%%%%%%%%%%%%%%%%%%%%%

	{\bf The quantum spin Hall (QSH) state was recently demonstrated in monolayers of the transition metal dichalcogenide 1T'-WTe$_2$ and is characterized by a band gap in the two-dimensional (2D) interior and helical one-dimensional (1D) edge states \cite{Crommie2017, Fei2017, Wu2018}.
		Inducing superconductivity in the helical edge states would result in a 1D topological superconductor, a highly sought-after state of matter \cite{Alicea2012}.
		In the present study, we use a novel dry-transfer flip technique to place atomically-thin layers of WTe$_2$ on a van der Waals superconductor, NbSe$_2$.
		Using scanning tunneling microscopy and spectroscopy (STM/STS), we demonstrate atomically clean surfaces and interfaces and the presence of a proximity-induced superconducting gap in the WTe$_2$ for thicknesses from a monolayer up to 7 crystalline layers.
		At the edge of the WTe$_2$ monolayer, we show that the superconducting gap coexists with the characteristic spectroscopic signature of the QSH edge state.
		Taken together, these observations provide conclusive evidence for proximity-induced superconductivity in the QSH edge state in WTe$_2$, a crucial step towards realizing 1D topological superconductivity and Majorana bound states in this van der Waals material platform.
		%   A quantum spin Hall (QSH) insulator is characterized by a band gap in its two-dimensional interior and helical one-dimensional (1D) edge states. Inducing superconductivity in the helical edge states results in a 1D topological superconductor, a highly sought-after state of matter at the core of many proposals for topological quantum computing. Using a novel %polymer-based 
		%   dry-transfer technique, we place atomically-thin layers of the QSH insulator 1T’$^{\prime}$-WTe$_2$ on a van der Waals superconductor, NbSe$_2$. Using scanning tunneling microscopy, we demonstrate the presence of a proximity-induced superconducting gap at the edge of a WTe$_2$ monolayer, which coexists with the characteristic spectroscopic signature of the QSH state.
		% This observation provides a clear pathway for realizing Majorana fermions in a van der Waals heterostructure.
	}
	
	Contemporary interest in topological superconductors has been driven by potential applications of their gapless boundary excitations, which are thought to be emergent Majorana quasiparticles with non-abelian statistics \cite{Kitaev2001,Fu2008,Sarma2015,Sato2017}.
	One path toward topological superconductivity is to realize an intrinsic spinless $p$-wave superconductor \cite{Maeno2012}.
	A powerful alternative is by using a conventional $s$-wave superconductor to induce Cooper pairing in topologically non-trivial states via the superconducting proximity effect, resulting in an effective $p$-wave pairing \cite{Fu2009}.
	This approach has recently been employed to engineer 2D topological superconductivity in epitaxial three-dimensional topological insulator films grown on a superconducting substrate \cite{Wang2012,Sun2016}, and 1D topological superconductivity by proximitizing a 2D QSH system in buried epitaxial semiconductor quantum wells \cite{Hart2014,Bocquillon2016}.
	While such demonstrations mark important milestones, there are clear advantages for exploring topological superconductivity in the van der Waals material platform. Using layered 2D materials allows the 2D QSH edge to be proximitized in vertical heterostructures, %from the vertical direction
	circumventing the length restrictions of lateral proximity-effect geometries.
	Furthermore, the surfaces and edges are readily available for surface probes, allowing detection and fundamental study of signatures of the topological superconducting state.
	Following recent theoretical predictions \cite{Qian2014}, an intrinsic QSH state was demonstrated in a monolayer (ML) of 1T'-WTe$_2$ \cite{Crommie2017,Fei2017,Jia2017,Peng2017,Wu2018,Shi2019}.
	WTe$_2$ is attractive for studying the QSH edge modes because it can be readily incorporated in van der Waals heterostructures and has shown quantized edge conductance up to \SI{100}{K} \cite{Wu2018}.
	Furthermore, ML~WTe$_2$ was recently also shown to host intrinsic superconducting behavior below \SI{\sim 1}{K} when electrostatically gated into the conduction band \cite{Fatemi2018,Sajadi2018}.
	%%%%%%%%%%%%%%%%%%%%%%%%%%%%%%%%%
	
	\begin{figure*}
		\includegraphics[width=0.9\textwidth]{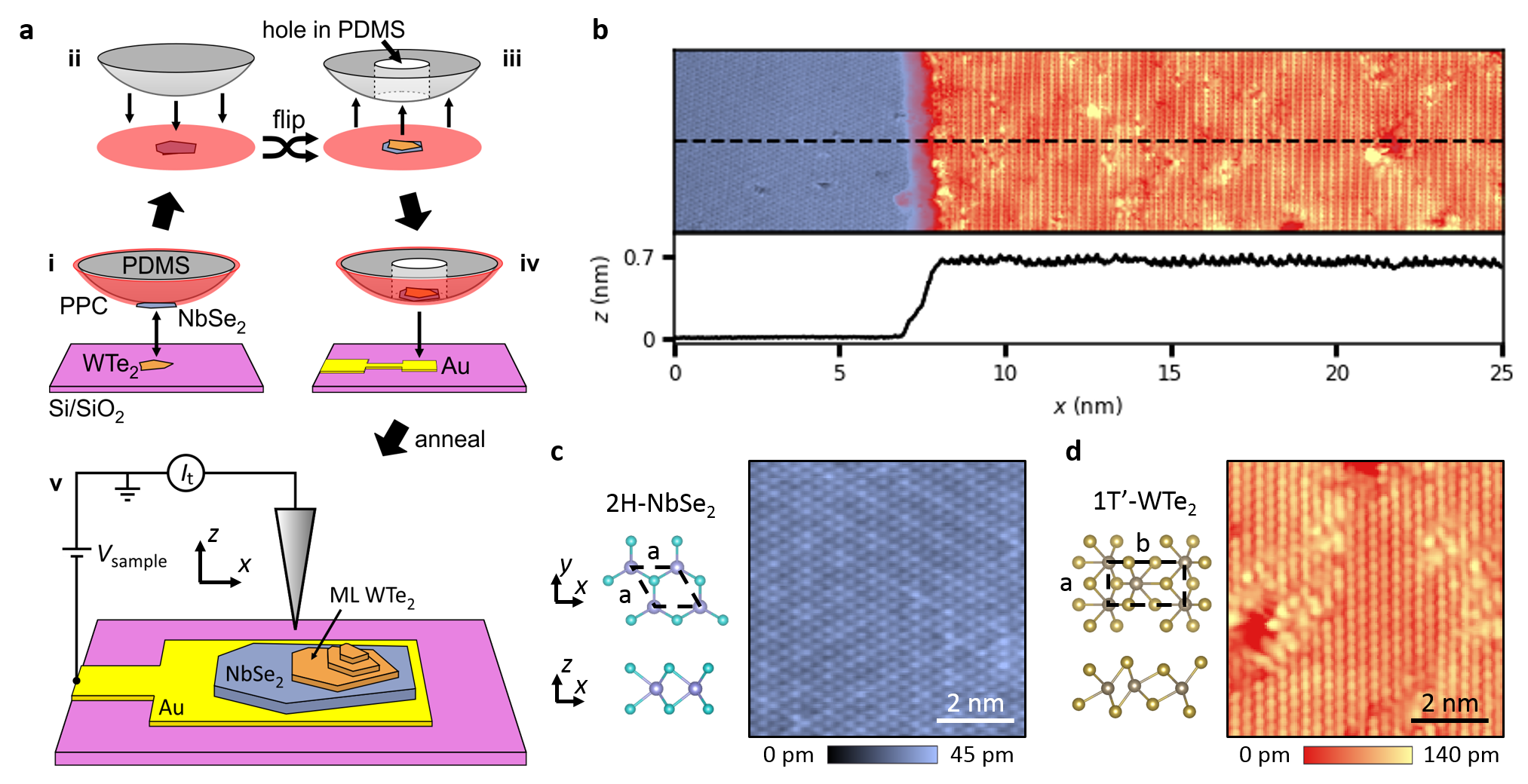}
		\caption{\label{Fig1} \textbf{Fabrication and morphology of WTe$_2$/NbSe$_2$ heterostructure.} 
			\textbf{(a)} Schematic of the sample fabrication and measurement setup.
			\textbf{(b)} STM topography and height profile across the edge of the monolayer WTe$_2$ flake ($V_{\rm sample} = 300\rm\,mV$ and $I_{\rm t} = 10\rm\,pA$). 
			\textbf{(c)} Atomic structures and atomically-resolved STM image of the NbSe$_2$ flake showing the $3\times3$ CDW ($V_{\rm sample} = 300\rm\,mV$ and $I_{\rm t} = 35\rm\,pA$). 
			\textbf{(d)} Atomic structures and atomically-resolved STM image of ML~WTe$_2$ ($V_{\rm sample} = 1\rm\,V$ and $I_{\rm t} = 55\rm\,pA$). 
			The topographies shown in (b), (c), and (d) are representative of the heterostructure over most of the area of the exfoliated flakes.
		}
	\end{figure*}   
	%%%%%%%%%%%%%%%%%%%%%%%%%%%%%%%%%%%
	
	In the present work, we study mechanically-exfoliated single- and few-layers of WTe$_2$ which have been transferred onto the van der Waals $s$-wave superconductor NbSe$_2$.
	We show that this approach induces a superconducting gap in the WTe$_2$ without the need for electrostatic doping and yields a critical temperature much higher than that of the intrinsic WTe$_2$ superconductivity, an experimental advantage which greatly facilitates studies of the interplay of superconductivity and the QSH edge modes.
	We employ scanning tunneling microscopy and spectroscopy (STM/STS) to investigate the proximity-induced superconducting gap as a function of temperature, magnetic field, and WTe$_2$ thickness.
	By spatially resolving the spectroscopic features of the WTe$_2$, we find that the superconducting gap coexists with the QSH signature at the ML~WTe$_2$ edge, demonstrating critical steps toward identifying 1D topological superconductivity in a van der Waals material system.
	
	We have developed a novel fabrication technique which enables the assembly and deterministic placement of van der Waals heterostructures in a glove box (Fig.~\ref{Fig1}a). Though similar methods have been used to fabricate 
	complex encapsulated mesoscale devices \cite{Zeng2018}, critically, our technique produces atomically-clean surfaces of air-sensitive materials suitable for high-resolution scanning probe measurements. In detail, WTe$_2$ and NbSe$_2$ are exfoliated from bulk materials and assembled using a PPC/PDMS stamp inside a nitrogen-filled glove box. Subsequently, the PPC film is peeled off, flipped upside down and put onto a second PDMS stamp which has a hole cut into it. This stamp is used to deterministically place the heterostructure onto a pre-patterned gold lead on a SiO$_2$ chip which is mounted and contacted to an STM sample plate. The PPC is then removed by annealing under vacuum conditions and the sample is transferred to the STM, all without intermediate air-exposure and without bringing the heterostructure surface into contact with any polymers or solvents, ensuring the surface cleanliness (for further details see Supplementary Materials). 
	Figure~\ref{Fig1}b shows an STM image of the resulting heterostructure where the WTe$_2$ ML edge and the underlying NbSe$_2$ are visible, showing atomically-clean surfaces on each material.
	The profile across the step edge shows a step height of \SI{\sim 7}{\AA} which corresponds to one WTe$_2$ layer \cite{Peng2017}, indicating an atomically-clean interface between the WTe$_2$ and NbSe$_2$. 
	In Fig.~\ref{Fig1}b a weak moir\'e pattern in the form of diagonal stripes can be seen on the ML~WTe$_2$ resulting from the superposition of the two different atomic lattices.
	The moir\'e pattern, analyzed in more detail in the Supplementary Materials, corresponds to a twist angle of $\approx3^{\circ}$ between the NbSe$_2$ hexagonal unit cell with lattice parameters $a=\SI{3.44}{\AA}$ and WTe$_2$ rectangular unit cell with lattice parameters $a=\SI{3.48}{\AA}$ and $b=\SI{6.28}{\AA}$ (Fig.~\ref{Fig1}c, d). 
	Atomically-resolved STM images of the NbSe$_2$~surface (Fig.~\ref{Fig1}c) show the well-known $3\times3$ charge density wave \cite{Wang2012}, indicating the pristine quality of the NbSe$_2$ flake. Atomically-resolved STM images of the WTe$_2$ (Fig.~\ref{Fig1}d) are characterized by vertical atomic rows parallel to the $a$-axis of the WTe$_2$ unit cell.
	
	%%%%%%%%%%%%%%%%%%%%%%%%%%%%%%%%%
	\begin{figure*}
		\includegraphics[width=0.7\textwidth]{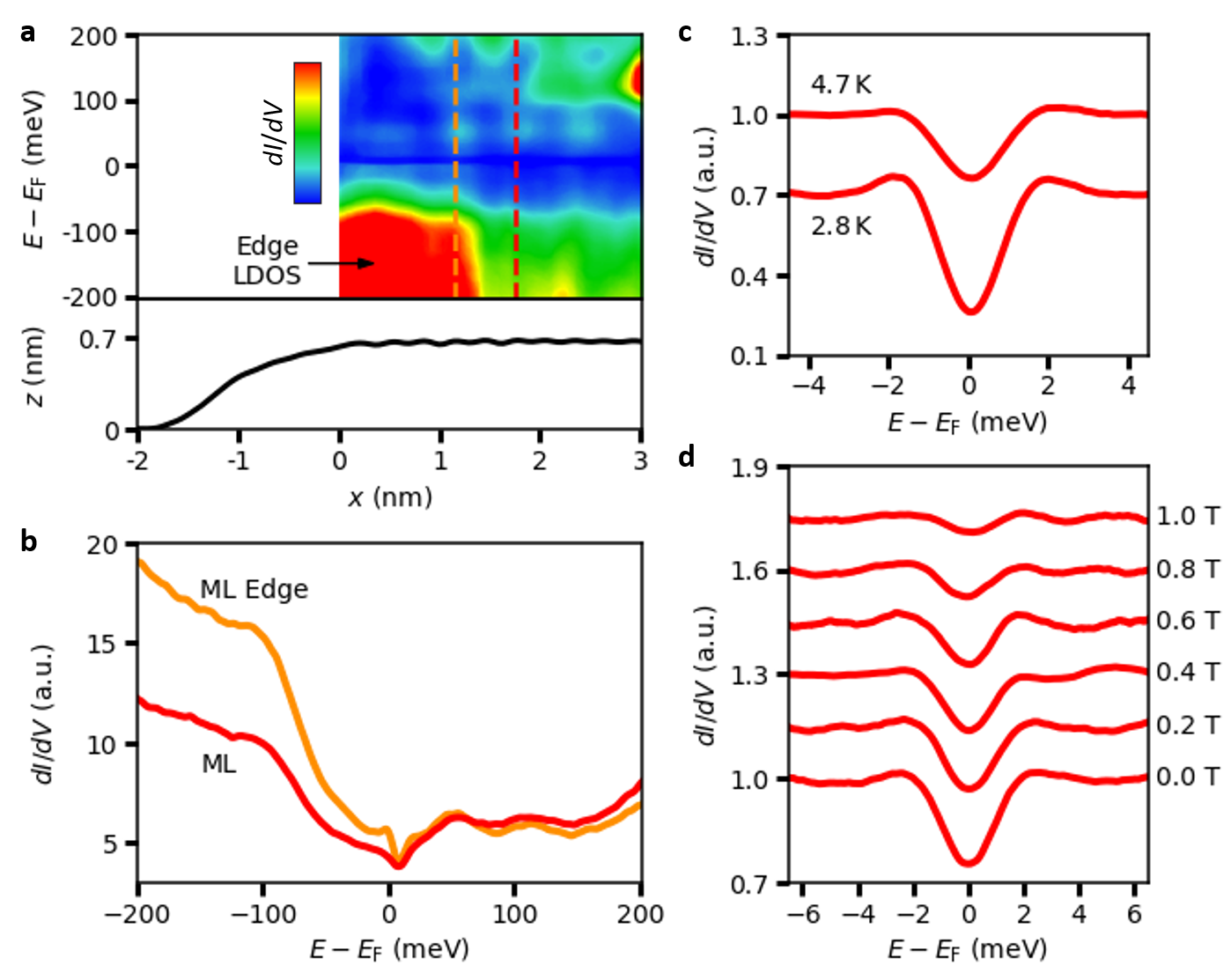}
		\caption{\label{Fig2} \textbf{Simultaneous presence of quantum spin Hall edge state and superconducting gap on monolayer WTe$_2$.} 
			\textbf{(a)} $dI/dV$ spectra taken along a line across the step edge of the WTe$_2$ flake (top), and corresponding height profile (bottom) 
			% 	($V_{\rm sample} = \SI{300}{mV}$ and $I_{\rm t} = \SI{400}{pA}$).
			\textbf{(b)} Spatially averaged $dI/dV$ spectra of monolayer WTe$_2$ showing a representative spectrum away from the monolayer edge (corresponding to the red dashed line in (a)) and increased density of states at the monolayer edge due to presence of the QSH edge state (corresponding to orange dashed line in (a)). Modulation amplitude $V_{\rm mod} = \SI{5}{mV}$. 
			\textbf{(c)} Small voltage range $dI/dV$ spectrum of monolayer WTe$_2$ at \SI{4.7}{K} and \SI{2.8}{K} showing superconducting gap-like features ($V_{\rm mod} = \SI{0.1}{mV}$). The \SI{2.8}{K} curve is offset for clarity. 
			\textbf{(d)} Magnetic field dependence of the small voltage range spectrum measured on the WTe$_2$ monolayer at \SI{4.7}{K}. The curves are offset for clarity.}
	\end{figure*}
	%%%%%%%%%%%%%%%%%%%%%%%%
	
	Turning now to spectroscopic analysis of these surfaces, Fig.~\ref{Fig2}a shows a series of $dI/dV$ spectra taken along a line perpendicular to the WTe$_2$ ML step edge (upper panel) and the corresponding height profile (lower panel).
	The $dI/dV$ spectra clearly show the presence of an increased local density of states (LDOS) near the WTe$_2$ step edge. 
	This feature was recently reported in STM/STS studies of ML films of WTe$_2$ grown on epitaxial graphene substrates \cite{Crommie2017,Jia2017}.
	Based on combined evidence from angle-resolved photoemission spectroscopy (ARPES) and STS in Ref.~\citenum{Crommie2017}, it was concluded that ML~WTe$_2$ has a band gap of \SI{56 \pm 14}{meV}, and that the increased LDOS at the ML~WTe$_2$ edge signifies the metallic QSH edge state.
	In our monolayer samples, produced via isolation from bulk crystals rather than molecular beam epitaxy, and on superconducting substrates rather than graphene, we observe the same spectroscopic features, which we attribute to the same QSH edge state.
	Figure~\ref{Fig2}b shows the averaged $dI/dV$ spectrum on the WTe$_2$ ML (red) and the ML edge (orange) at the corresponding positions indicated in Fig.~\ref{Fig2}a.
	The spectroscopic signature of the QSH edge state is evident primarily in the valence band but, importantly, the edge state also crosses the band gap.
	Following the interpretation of Ref.~\citenum{Crommie2017}, the increases in the $dI/dV$ signal at $E-E_{\mathrm{F}}\approx-\SI{50}{meV}$ and $E-E_{\mathrm{F}}\approx\SI{15}{meV}$ correspond to the onset of the WTe$_2$ valence and conduction band, respectively, locating $E_F$ in the ML WTe$_2$ band gap. 
	A non-zero $dI/dV$ signal in the band gap away from the step edge was proposed to be due to defect states and substrate effects \cite{Crommie2017}.
	In addition, tip-induced band bending may play a role in introducing spectral weight in the WTe$_2$ band gap (see Supplementary Materials). By comparing the positions of the observed spectral features to epitaxially-grown WTe$_2$ on graphene \cite{Crommie2017,Jia2017} and exfoliated WTe$_2$ \cite{Cucchi2018}, we conclude that there is no significant charge transfer from the NbSe$_2$ to the WTe$_2$.
	This observation is further supported by our density functional theory (DFT) calculations of the ML~WTe$_2$/NbSe$_2$ heterostructure, which show only minimal modifications of the WTe$_2$ electronic structure compared to a freestanding WTe$_2$ ML (see Supplementary Materials).
	
\begin{figure*}
		\includegraphics[width=0.8\textwidth]{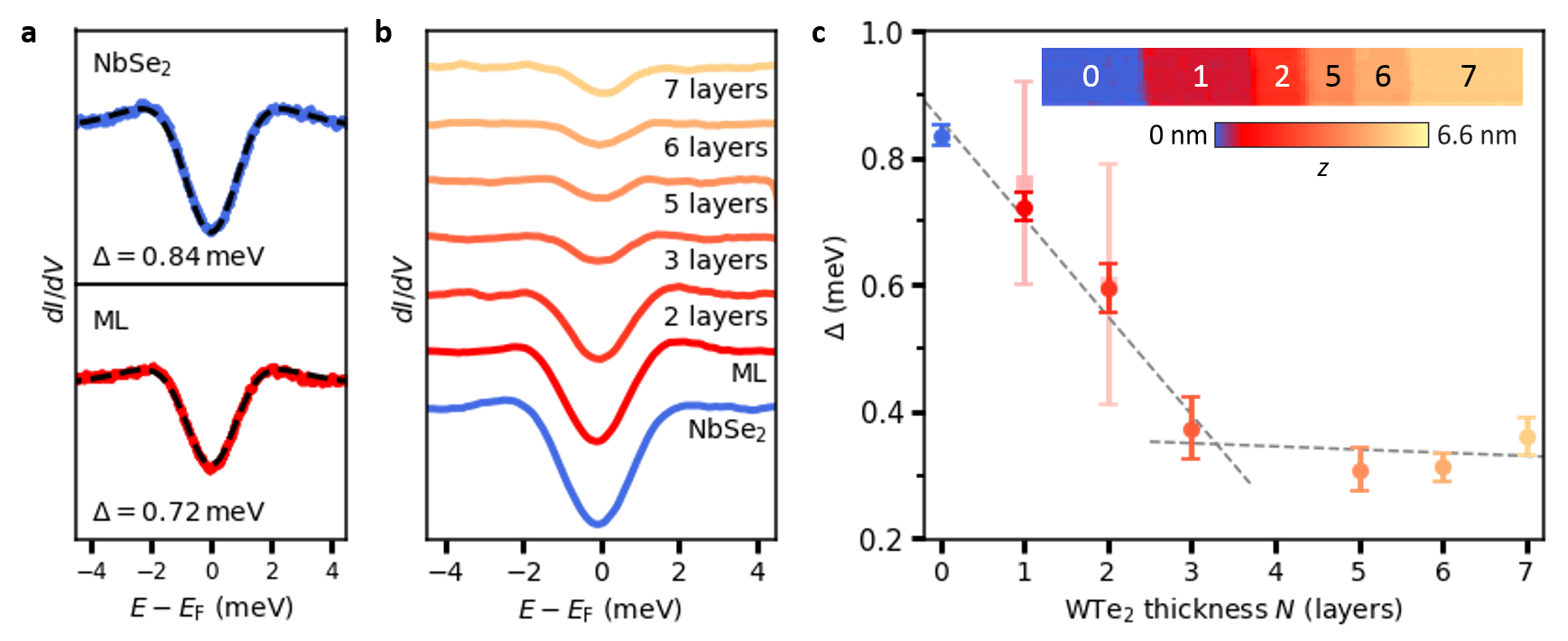}
		\caption{\label{Fig3}
			\textbf{Evolution of the superconducting gap with WTe$_2$ thickness at 4.7 K.} 
			\textbf{(a)} Fits of the BCS model to the superconducting gap spectra measured on NbSe$_2$ and monolayer WTe$_2$.
			\textbf{(b)} Measurement of the superconducting gap spectrum for WTe$_2$ layer thicknesses up to 7 layers. 
			\textbf{(c)} (Filled circles) WTe$_2$ thickness dependence of the superconducting gap size obtained from fitting the spectra in (b) with the BCS gap equation. (Filled squares) Fits of the monolayer and bilayer spectrum with a more detailed model which includes partial tunneling into the NbSe$_2$ substrate.
			The dashed lines are guides to the eye and indicate two different regimes in which $\Delta$ decreases more rapidly for $N<3$ and more gradually for $N\geq 3$. The inset shows a large-scale topography image of the WTe$_2$, where terraces of different WTe$_2$ thicknesses are observed. Scan size: $200\rm\,nm\times14\,nm$. The corresponding number of WTe$_2$ layers $N$ is indicated for each terrace, where $N=0$ is the bare NbSe$_2$.
		}
	\end{figure*}

	Measurements of the ML~WTe$_2$ $dI/dV$ spectrum over a smaller voltage range (Fig.~\ref{Fig2}c), reveal a new feature: a superconducting gap-like feature characterized by a dip in the $dI/dV$ signal at the Fermi energy, with peaks on either side of the gap.
	Comparison of measurements at \SI{4.7}{K} and \SI{2.8}{K} show that the gap deepens and the peaks sharpen at lower temperature. 
	The evolution of the superconducting gap-like feature under application of a surface-normal magnetic field at \SI{4.7}{K} (Fig.~\ref{Fig2}d) shows that with increasing magnetic field, the gap becomes less deep and the peaks flatten out until the gap features have nearly vanished at \SI{1}{T}. 
	We find that a fit of the Bardeen-–Cooper-–Schrieffer (BCS) model describes both the monolayer WTe$_2$ and the NbSe$_2$ data well (Fig.~\ref{Fig3}a).
	For NbSe$_2$, the fit results in a superconducting gap of $\Delta_{\rm NbSe_2}=\SI{0.84 \pm 0.01}{meV}$, while for the WTe$_2$ we find $\Dbulk=\SI{0.72\pm0.02}{meV}$.
	In addition to following the trend of a superconducting gap with applied magnetic field, the vanishing of the gap near \SI{1}{T} is similar to the Ginzburg–-Landau estimate for the upper critical field
	of bulk NbSe$_2$ \cite{Garoche1976}.
	We conclude that the gap feature observed on the ML~WTe$_2$ is indeed a superconducting gap.
	
	%%%%%%%%%%%%%%%%%%%%%%%%%%%%%%%%%%%%
	%%%%%%%%%%%%%%%%%%

\begin{figure*}
		\includegraphics[width=0.8\textwidth]{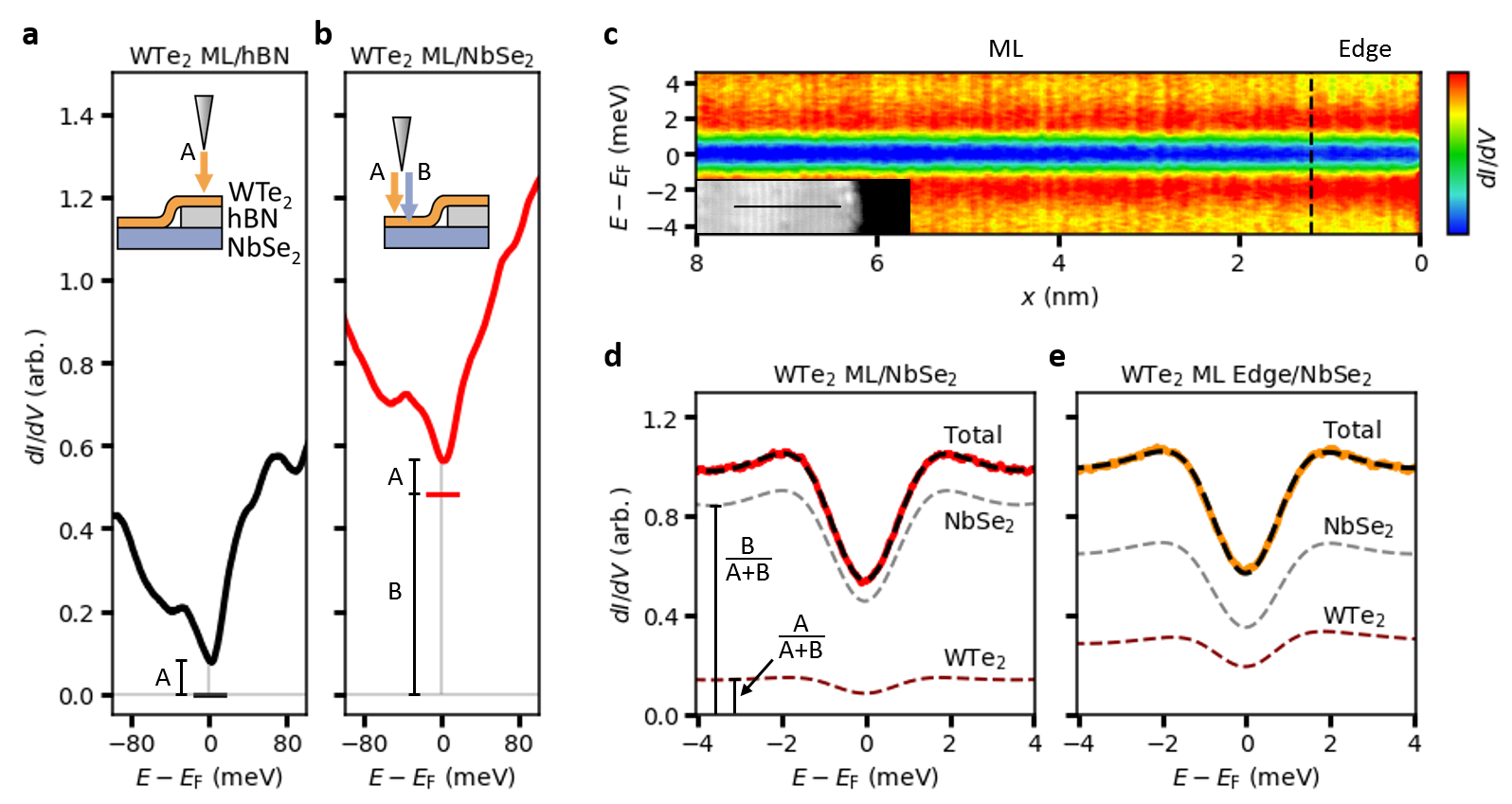}
		\caption{\label{Fig4} 
			\textbf{Proximity-induced superconducting gap in the quantum spin Hall edge state of monolayer WTe$_2$ at 2.8 K.}  \textbf{(a)} Tunneling spectrum of WTe$_2$ on hexagonal boron nitride (hBN) and 
			\textbf{(b)} spectrum of the \textit{same} WTe$_2$ flake on NbSe$_2$ (for optical micrograph of the heterostructure see Supplementary Materials). The tunneling contributions of the WTe$_2$ and NbSe$_2$ are denoted as $A$ and $B$, respectively. 
			\textbf{(c)} SC gap spectra measured along a line perpendicular to the edge of the WTe$_2$. The inset shows the topography and line along which the spectra were taken. Scan size: \SI{16}{nm} $\times$ \SI{4}{ nm}.
			\textbf{(d)} Fitting of representative ML WTe$_2$/NbSe$_2$ tunneling spectrum with fractional WTe$_2$ and NbSe$_2$ contributions determined from (a) and (b). 
			The grey and maroon dashed lines indicate the relative contributions $(dI/dV)_{\rm{NbSe_2}}$ and $(dI/dV)_{\rm{WTe_2}}$.
			\textbf{(e)} Fitting of the ML edge WTe$_2$/NbSe$_2$ tunneling spectrum with fractional WTe$_2$ and NbSe$_2$ contributions determined as for (d) but with the increased DOS of the edge state.
		}
	\end{figure*}

	In order to confirm the proximity-induced nature of the observed superconducting gap on the WTe$_2$, we explore its evolution as a function of WTe$_2$ thickness.
	The exfoliation procedure naturally produces terraces of varying thickness in our samples, enabling thickness-dependent gap measurement within a single sample.
	Figure~\ref{Fig3}b shows the superconducting gap measured on terraces of different numbers of WTe$_2$ layers $N$, revealing that the gap decreases with increasing $N$, 
	as expected for decaying superconducting correlations near the boundary of a superconducting--metal interface \cite{Huang2018}.
	To quantify this behavior, we fit the BCS model to each of the spectra in Fig.~\ref{Fig3}b and plot the extracted gap sizes as filled circles in Fig.~\ref{Fig3}c.
	In the thick limit ($N\geq$3), we find that observed behavior shows excellent agreement with transport measurements of proximity-induced superconductivity in bulk WTe$_2$ flakes \cite{Huang2018,Li2018}, extending the previous studies to the ultra-thin limit (see Supplementary Materials).
	For $N < 3$, we observe a more rapid decrease of the extracted gap as function of $N$ which may be explained by the strong variation of the electronic structure of the WTe$_2$ in this thickness range, resulting in a larger mismatch of the WTe$_2$ and NbSe$_2$ Fermi surfaces and therefore a stronger dependence of the induced gap on $N$ \cite{Reeg2016}. 
	For monolayer and bilayer WTe$_2$, we also consider the possibility of tunneling spectra being a superposition of tunneling into \wte and into \nbse (Fig.~\ref{Fig4}b inset).
	To isolate the respective contributions, we performed a control experiment on a second sample, in which we placed a \SI{\sim 20}{nm} thick layer of insulating hBN between the WTe$_2$ and NbSe$_2$ to locally decouple the WTe$_2$ (Fig.~\ref{Fig4}a, b). By comparing the measured $dI/dV$ signal at the Fermi energy of spectra taken on WTe$_2$/hBN ($A$) and WTe$_2$/NbSe$_2$ ($B$), we find that the fractional contribution of tunneling into ML WTe$_2$ on the NbSe$_2$ is $f_{\rm{WTe_2}} \equiv A/(A+B)=0.14 \pm 0.04$. The NbSe$_2$ contribution is correspondingly $f_{\rm{NbSe_2}} = B/(A+B)=0.86 \pm 0.04$, i.e. the majority of the tunneling current.  
	Knowing the relative WTe$_2$ contribution allows us to perform a more detailed analysis of the superconducting ML WTe$_2$/\nbse spectrum (Fig.~\ref{Fig4}d) by fitting a superposition of WTe$_2$ and NbSe$_2$ BCS spectra with the same fractional contributions, i.e. $(dI/dV)_{\rm Total} = f_{\rm{WTe_2}}\cdot (dI/dV)_{\rm{WTe_2}}+f_{\rm{NbSe_2}}\cdot (dI/dV)_{\rm{NbSe_2}}$, using a BCS form for each $dI/dV$ (details in Supplementary Materials).  
	The resulting proximity-induced gap sizes from the more detailed analysis are $\Dbulk = \SI{0.76\pm0.16}{meV}$ at \SI{4.7}{K} and \SI{0.83\pm0.08}{meV} at \SI{2.8}{K}. For bilayer WTe$_2$ on NbSe$_2$, using a similar procedure, we find an induced gap of $\DbulkBL = \SI{0.60\pm0.19}{meV}$.  In Fig.~\ref{Fig3}c we also plot the \SI{4.7}{K} WTe$_2$ proximity gaps found from this more detailed fitting and find no significant deviation from those previously determined by fitting the single-gap BCS theory.
	This observation is in qualitative agreement with theory which predicts the proximity-induced gap to approach that of the NbSe$_2$ as the WTe$_2$ layer thickness goes to zero \cite{Huang2018}.
	
	%%%%%%%%%%%%%%%%%%%%%%%%%%%%%%%%%%%%
	%%%%%%%%%%%%%%%%%%%%%%%%%%%%%%%%%%%%
	
	Finally, we consider the lateral variation of the superconducting gap from within the ML WTe$_2$ to the region occupied by the edge state. Figure~\ref{Fig4}c shows $dI/dV$ spectra taken at \SI{2.8}{K} along a line approaching the physical edge of the WTe$_2$ monolayer, similar to that shown in Fig.~\ref{Fig2}a but over a smaller voltage range.  We find that the superconducting gap is present throughout the WTe$_2$ monolayer with only slight changes in the gap width and depth.  
	It is apparent that a superconducting gap is present in the region in which the QSH edge state is observed in Fig.~\ref{Fig2}a (indicated by the dashed line in Fig.~\ref{Fig4}c).
	To determine the fractional contribution of the WTe$_2$ edge spectrum, in Fig.~\ref{Fig4}e we perform a similar fit as we did for the spectrum away from the edge (Fig.~\ref{Fig4}d), using the same absolute NbSe$_2$ background as in Fig.~\ref{Fig4}d, but the larger conductance for of the WTe$_2$, corresponding to the larger tunneling conductance into the edge state at the Fermi energy (Fig.~\ref{Fig2}b). The resulting relative contribution of the edge state is $f_{\rm{WTe_2}}=0.33 \pm 0.06$ and the extracted gap size is $\Dedge = \SI{0.75 \pm 0.08}{meV}$.

	The observation of a superconducting gap in the edge state of monolayer 1T'-WTe$_2$ provides strong evidence that we have created a 1D topological superconductor in a van der Waals heterostructure by proximity-induced superconductivity in the quantum spin Hall edge state.
	The topological nature of a superconducting QSH edge state could be explicitly demonstrated in an STM measurement by creating a boundary with a portion of the same QSH edge state in which a topologically-trivial gap has been opened \cite{Alicea2012}.
	This would localize Majorana zero modes at the boundary, which can be identified as a zero-bias conductance peak within the superconducting gap \cite{Jaeck2019}.
	Creating such a boundary is straightforward in the van der Waals material platform, e.g., by integrating a van der Waals magnetic insulator into the heterostructure shown in Fig.~\ref{Fig1}a to open a local Zeeman gap.
	Our work establishes the groundwork for such an experiment with a clear path toward the realization of Majorana quasiparticles.
	In addition, the method of sample preparation outlined in this work may be easily adapted to numerous experiments involving surface-probe studies or air-sensitive materials.
	
	% \section*{References and Notes}
	% \begingroup
	% \renewcommand{\section}[2]{}%
	% \bibliographystyle{naturemag}
	% \bibliography{Refs.bib}
	
	%\section*{Acknowledgments:}
	\paragraph{Acknowledgments.}
	The authors acknowledge Di Xiao, David Cobden and Xiaodong Xu for helpful discussions and Nicholas Speeney and Nicolas Iskos for assistance in the lab.
	B.M.H. was supported by the Department of Energy under the Early Career award program ({DE-SC0018115}). Crystal growth and characterization at ORNL was supported by the US Department of Energy, Office of Science, Basic Energy Sciences, Division of Materials Sciences and Engineering. The authors thank the Pennsylvania State University Two-Dimensional Crystal Consortium - Materials Innovation Platform (2DCC-MIP) which is supported by NSF DMR-1539916 for supplying further 2D materials. F.L. and D.W. were supported by the NSF DMR-1809145 for the STM measurements. The authors gratefully acknowledge NSF DMR-1626099 for acquisition of the STM instrument. S.C.d.l.B. was supported by the Department of Energy ({DE-SC0018115}) for fabrication of proximity-effect van der Waals heterostructures. DFT Calculations were supported by the Department of Energy under grant {DE-SC0014506}. 
	
	\vspace{3mm}
	
	\paragraph{Author contributions.}
	F.L., D.W., R.M.F. and B.M.H. designed the experiment.  F.L. and D.W. acquired the experimental data and F.L., D.W., and R.M.F. analyzed the data. F.L., D.W. and S.C.d.l.B. fabricated the samples. F.L., D.W., S.C.d.l.B., R.M.F. and B.M.H. wrote the manuscript, and all authors commented on the manuscript. J.Y. grew the WTe$_2$ crystals. D.M. provided other van der Waals crystals used in this study. M.W. performed DFT calculations. R.M.F. and B.M.H. supervised the project.
	%\paragraph{Competing interests} The authors declare no competing interest.

	% \endgroup

	\newpage

	%%%%%%%%%%%%%Supplementary Material%%%%%%%%%%%%%%%%%%

	\clearpage
	% \appendix
	% \documentclass[12pt]{article}
	%\usepackage[T1]{fontenc}
	% \usepackage{times}
	
	% New names of stuff for Supp info
	% \setcounter{figure}{0}
	% \setcounter{equation}{0}
	% \setcounter{mybibstartvalue}{27}
	% \xpatchcmd{\thebibliography}{%
	%   \usecounter{enumiv}%
	% }{%
	%   \usecounter{enumiv}%
	%   \setcounter{enumiv}{\value{mybibstartvalue}}%
	% }{}{}

	% \renewcommand\thesection{S\arabic{section}}
	% \renewcommand\thesubsection{S\thesection.\arabic{subsection}}
	% % \renewcommand{\thesection}{\Roman{section}} 
	% % \renewcommand{\thesubsection}{\Roman{subsection}}
	% \renewcommand{\thefigure}{S\arabic{figure}}
	% \captionsetup[figure]{name={\bf{Fig.}}}
	% \captionsetup[table]{name={\bf{Table}}}
	
	% \renewcommand{\theequation}{S\arabic{equation}}
	% \renewcommand{\thetable}{S\arabic{table}}

	\author
	{Felix L\"upke,$^{1\ast}$ Dacen Waters,$^{1\ast}$ Sergio C. de la Barrera,$^{1}$ Michael Widom,$^{1}$\\ David G. Mandrus,$^{2,3,4}$ Jiaqiang Yan,$^{2}$ Randall M. Feenstra,$^{1}$ Benjamin M. Hunt$^{1\dagger}$\\
		\\
		\normalsize{$^{1}$Department of Physics, Carnegie Mellon University, Pittsburgh, PA 15213, USA}\\
		\normalsize{$^{2}$Materials Science and Technology Division, Oak Ridge National Laboratory, Oak Ridge, TN 37831, USA}\\
		\normalsize{$^{3}$Department of Materials Science and Engineering, University of Tennessee, Knoxville, TN 37996, USA}\\
		\normalsize{$^{4}$Department of Physics and Astronomy, University of Tennessee, Knoxville, TN 37996, USA}\\
		\\
		\normalsize{$^\ast$ These authors contributed equally}\\
		\normalsize{$^\dagger$Corresponding author. E-mail:  bmhunt@andrew.cmu.edu}
	}
	% \date{}
	
	%%%%%%%%%%%%%%%%%%%%%%%% SUPPLEMENTARY INFO
	\onecolumngrid % comment to retain two-column format
	\cleardoublepage
	\setcounter{figure}{0}
	\setcounter{equation}{0}
	\renewcommand\thesection{S\arabic{section}}
	\renewcommand\thesubsection{S\thesection.\arabic{subsection}}
	\renewcommand{\thefigure}{S\arabic{figure}}
	\renewcommand{\theequation}{S\arabic{equation}}

	\makeatletter
	\renewcommand\@biblabel[1]{[S#1]} % this replaces the natbib command 'bibnumfmt' below, which was not working
	\makeatother
	\renewcommand{\citenumfont}{S}
	
	\begin{center}
		
		{\textbf{\large Supplementary Materials: \\Proximity-induced superconducting gap in the quantum spin Hall edge state of monolayer WTe$_2$\\}}
		\vspace{0.5cm}
		{Felix L\"upke,$^{1\ast}$ Dacen Waters,$^{1\ast}$ Sergio C. de la Barrera,$^{1}$ Michael Widom,$^{1}$\\ David G. Mandrus,$^{2,3,4}$ Jiaqiang Yan,$^{2}$ Randall M. Feenstra,$^{1}$ and Benjamin M. Hunt$^{1\dagger}$}
		\vspace{0.3cm}
		
		\normalsize{$^{1}$\textit{Department of Physics, Carnegie Mellon University, Pittsburgh, PA 15213, USA}}\\
		\normalsize{$^{2}$\textit{Materials Science and Technology Division, Oak Ridge National Laboratory, Oak Ridge, TN 37831, USA}}\\
		\normalsize{$^{3}$\textit{Department of Materials Science and Engineering, University of Tennessee, Knoxville, TN 37996, USA}}\\
		\normalsize{$^{4}$\textit{Department of Physics and Astronomy, University of Tennessee, Knoxville, TN 37996, USA}}\\
		\normalsize{$^\ast$ These authors contributed equally}\\
		\normalsize{$^\dagger$E-mail: bmhunt@andrew.cmu.edu}
		\vspace{0.5cm}
		
	\end{center}
	
	\section{Sample fabrication}

\begin{figure*}[h]
		\centering
		\includegraphics[width=12cm]{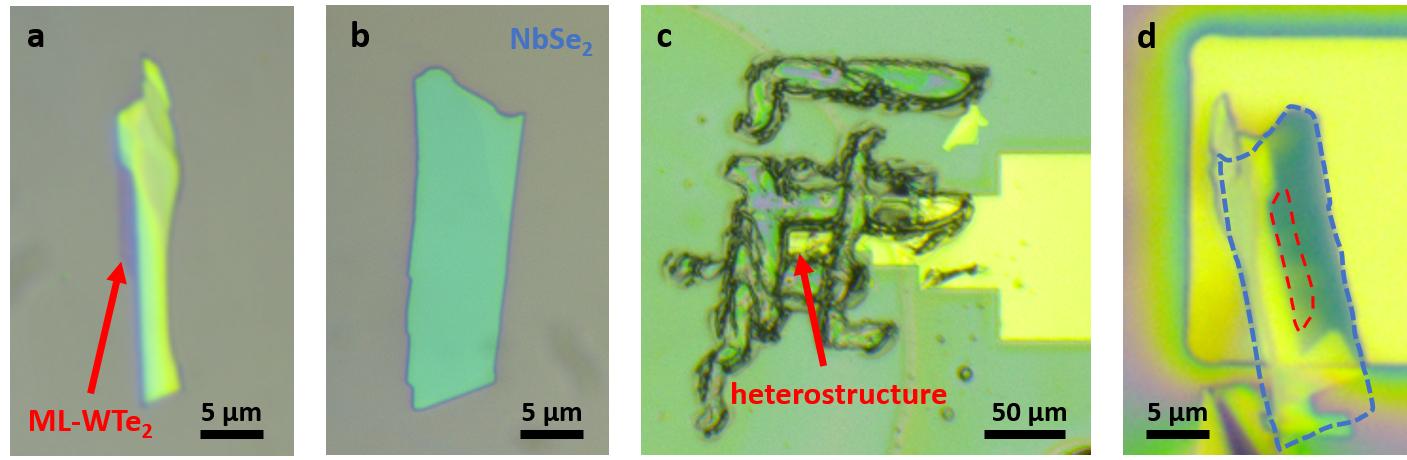}
		\caption{\label{FigS1} {\bf Fabrication of the WTe$_2$/NbSe$_2$ heterostructure.} \textbf{(a), (b)} Optical image of the WTe$_2$ and NbSe$_2$ flake after exfoliation onto SiO$_2$. 
			\textbf{(c)} Optical image after putting down the flipped heterostructure onto the gold lead and removing some of the PPC surrounding the flake using a sharp needle.
			\textbf{(d)} Optical image showing the heterostructure on the gold lead after vacuum annealing. Red and blue dashed lines indicate the outlines of the WTe$_2$ monolayer and NbSe$_2$, respectively.
		}
	\end{figure*}

	The newly developed sample fabrication method consists of the following steps: 
	\begin{enumerate}
		\item Mechanically exfoliate NbSe$_2$ and WTe$_2$ onto O$_2$-plasma-cleaned SiO$_2$/Si wafer in a nitrogen-filled glove box (Fig.~\ref{FigS1}a and b). 
		\item Using a PPC/PDMS droplet transfer slide \cite{Wang2013-SI}, pick up the NbSe$_2$ flake, then pick up the WTe$_2$ flake. The transfer stage temperature is $\approx40^{\circ}$C for both pick ups.
		\item Peel off the PPC layer from the PDMS, flip it upside down and put it onto a second transfer slide which has a PDMS droplet on it into which a hole was cut such that the heterostructure sits in the center of the hole and does not touch the PDMS.
		\item Place the heterostructure onto pre-evaporated gold leads (Au (\SI{50}{nm})/Pd (\SI{30}{nm})/Cr (\SI{10}{nm})) on an SiO$_2$/Si chip which is mounted on an STM sample plate and wire bonded to contacts sitting on the sample plate. Release the PPC by heating it to $100^{\circ}$C.
		\item With the sample still at $100^{\circ}$C, remove some of the PPC surrounding the heterostructure using a sharp needle (Fig. ~\ref{FigS1}c). This step prevents the heterostructure from floating off of the gold lead in the next step.
		\item Transfer the sample from the glovebox into vacuum (without exposing it to air), and anneal it at $250^{\circ}$C for $8\rm\,h$ ($p\leq1\cdot10^{-5}\rm\,mBar$) to remove the PPC. We perform this step using a tube furnace which has a gate valve attached to it and is vented with nitrogen. Subsequently, transfer the sample to the STM ultra-high vacuum chamber and perform a final annealing step at $250^{\circ}$C for $\SI{\sim10}{mins}$ ($p\leq1\cdot10^{-8}\rm\,mBar$) before introducing the sample into the STM.
	\end{enumerate}
	Using this technique, in contrast to previously reported dry-transfer techniques, the final top surface of the assembled van der Waals heterostructure is never in contact with any polymer or solvent. We find this to be crucial to achieve atomically-clean surfaces.
	Furthermore, sample fabrication is performed entirely in an inert gas/vacuum environment. This allows to study the pristine surfaces of highly air-sensitive materials, the only constraint being that the materials do not degrade when annealed at $250^{\circ}\rm C$ in vacuum. 
	
	For the samples presented in the main text, WTe$_2$ flakes were transferred onto \SI{\sim20}{nm} thick NbSe$_2$ flakes. 
	At this thickness, the electronic properties of the NbSe$_2$ are bulk-like and the critical temperature below which the NbSe$_2$ becomes superconducting is $T_{\rm c}\approx\SI{7}{K}$ \cite{Khestanova2018-SI}. Figure \ref{FigS1} shows optical images of the sample studied in the main text, except in Fig.~\ref{Fig4}a and b.
	
	\section{Scanning tunneling measurements}
	The STM tip is approached to the WTe$_2$/NbSe$_2$ heterostructure using a capacitive technique adapted from Ref. \citenum{Li2011-SI}. The commercial CreaTec STM helium bath temperature is $4.2\rm\,K$ with the ability of intermittently reaching $\sim 1\rm\,K$ by pumping on the cryostat. The resulting temperatures of the STM base plate are $4.7\rm\,K$ and $2.8\rm\,K$, respectively, due to vibration isolation and optical access. The STM is equipped with an electrochemically-etched tungsten tip which was cleaned by indentation into gold prior to and in between measurements. The lock-in frequency was set to $f=925\rm\,Hz$ in all $dI/dV$ measurements. All superconducting gap measurements were performed at $V_{\rm sample} = 5\rm\,mV$ with $V_{\rm mod} = 100\rm\,\murm V$ peak-to-peak and $I_{\rm t} = 100\rm\,pA$, except in Fig.~\ref{Fig2}e where $V_{\rm sample} = 10\rm\,mV$. The spectra in Fig.~\ref{Fig2}a and b were acquired using $V_{\rm sample} = \SI{300}{mV}$, $I_{\rm t} = \SI{400}{pA}$ and $V_{\rm mod} = 5\rm\,\rm mV$. 
	In Fig.~\ref{Fig4}, tunneling parameters are: $V_{\rm sample} = 300\rm\,mV$, $I_{\rm t} = 100\rm\,pA$ and $V_{\rm mod} = 5\rm\,mV$ in (a) and $V_{\rm sample} = 300\rm\,mV$, $I_{\rm t} = 110\rm\,pA$ and $V_{\rm mod} = 10\rm\,mV$ in (b). For quantitative comparison, the spectra in Fig.~\ref{Fig4}a and b and Fig.~\ref{Fig:WTe2-hBN}c were normalized to $I_{\rm t}$ and $V_{\rm mod}$, respectively.

	\section{Moir\'e effects}
	
\begin{figure*}[h]
		\centering
		\includegraphics[width=14cm]{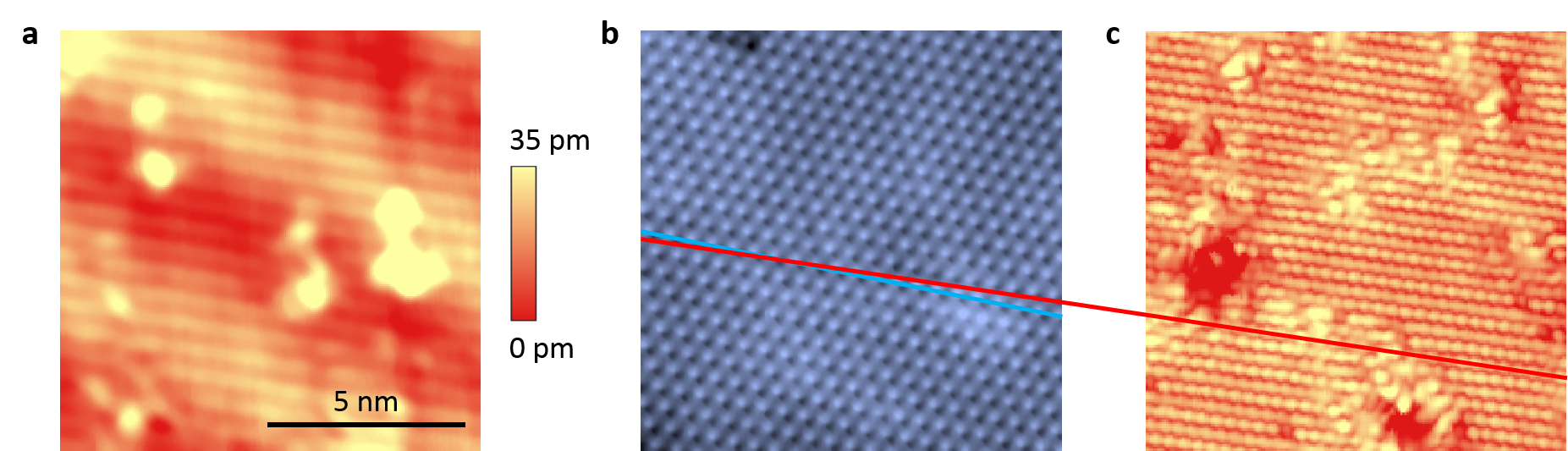}
		\caption{\label{FigS2}
			{\bf Observed moir\'e pattern.} \textbf{(a)} Topography image of the ML WTe$_2$/NbSe$_2$ heterostructure, showing the moir\'e periodicity and WTe$_2$ atomic rows ($V_{\rm sample} = 1\rm\,V$ and $I_{\rm t} = 50\rm\,pA$).
			\textbf{(b)} Atomic resolution image of NbSe$_2$ right below the step edge of the WTe$_2$. The blue line indicates the orientation of the NbSe$_2$ lattice vector.
			\textbf{(c)} Atomic resolution image of WTe$_2$ right next to the step edge. The red line indicates the orientation of the WTe$_2$ $a$-lattice vector.
			This misorientation with respect to the NbSe$_2$ is $\sim3^{\circ}$.
		}
	\end{figure*}

\begin{figure*}[ht!]
		\centering
		\includegraphics[width=13cm]{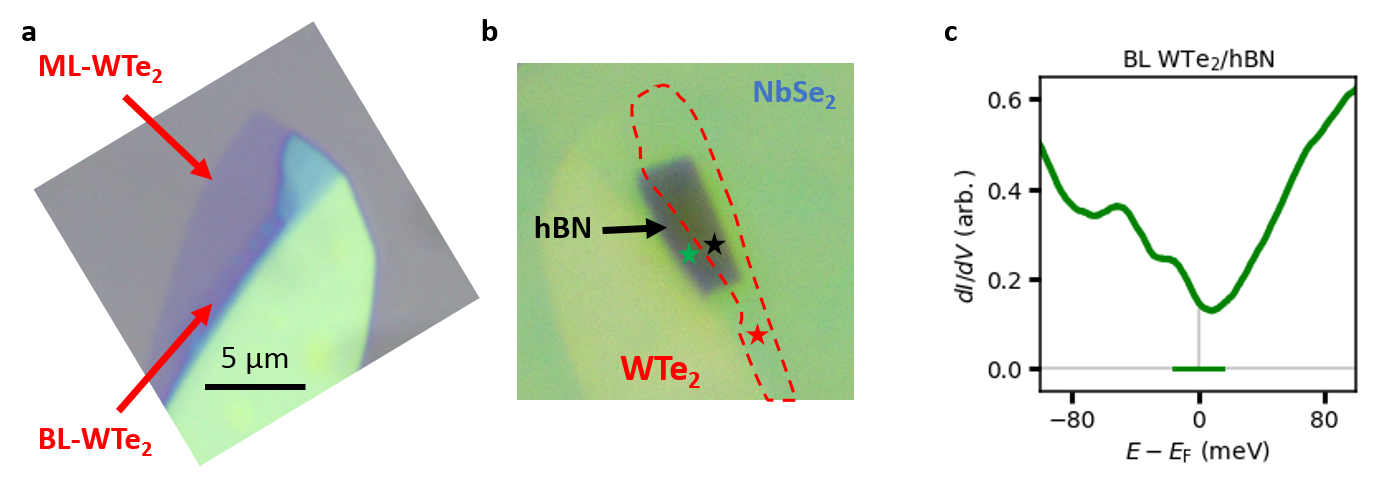}
		\caption{\label{Fig:WTe2-hBN}
			{\bf WTe$_2$/hBN/NbSe$_2$ heterostructure}. \textbf{(a)} Optical image of the WTe$_2$ flake after exfoliation, with indicated WTe$_2$ monolayer and bilayer regions.
			\textbf{(b)} Optical image of the assembled WTe$_2$/hBN/NbSe$_2$ heterostructure. %There is a small area of bilayer (BL) immediate adjacent to the ML (indicated by the green star). 
			The hBN thickness is $\sim\SI{20}{nm}$ and the NbSe$_2$ thickness is $\sim\SI{100}{nm}$. The red, black and green stars indicate the positions at which spectra of ML~WTe$_2$/hBN (Fig.~\ref{Fig4}a), ML~WTe$_2$/NbSe$_2$ (Fig.~\ref{Fig4}b) and BL~WTe$_2$/hBN \textbf{(c)} were taken, respectively. 
		}
	\end{figure*}

	Figure~\ref{FigS2}a shows an STM image where both the moir\'e periodicity and the atomic rows of the ML WTe$_2$/NbSe$_2$ heterostructure can be seen. Analyzing atomic resolution images of the WTe$_2$ and NbSe$_2$ close to the WTe$_2$ step edge, we find a small rotational misalignment of $\sim3^{\circ}$ between the WTe$_2$ $a$-axis and the NbSe$_2$ (Fig.~\ref{FigS2}b and c). The resulting moir\'e period can be approximated as
	\begin{equation}
	L = \frac{a}{\sqrt{\Theta^2+\varepsilon^2}}
	\end{equation}
	where $a=3.48$ \AA\ is the shorter WTe$_2$ lattice constant, $\varepsilon$ is its relative lattice mismatch with respect to the NbSe$_2$ lattice constant ($\sim 1.1\%$) and $\Theta\approx3^{\circ}$ is the rotational angle between the two materials \cite{Jung2014-SI}.
	The resulting moir\'e period is $L\approx\SI{6.5}{nm}$ which is in good agreement with the periodicity observed in the STM image. Note that the moir\'e period in direction of the WTe$_2$ $b$ vector is negligibly small because the rotational misalignment of it with respect to the second NbSe$_2$ lattice vector is large.
	
	%%%%%%%%%%%%%%%%%%%%%%%%%%%%%%%%%%%%%%%%%%%%%%%%%%%%%%%%%%%%%%%%%%
	%%%%%%%%%%%%%%%%%%%%%%%%%%%%%%%%%%%%%%%%%%%%%%%%%%%%%%%%%%%%%%%%%%

	%%%%%%%%%%%%%%%%%%%%%%%%%%%%%%%%%%%%%%%%%%%%%%%%%%%%%%%%%%%%%%%%%%
	%%%%%%%%%%%%%%%%%%%%%%%%%%%%%%%%%%%%%%%%%%%%%%%%%%%%%%%%%%%%%%%%%%
	
	\section{Band gap in ML WTe$_2$}
	While ARPES \cite{Crommie2017-SI} and transport measurements \cite{Wu2018-SI} show indications of a band gap in the electronic structure of ML WTe$_2$, so far no hard gap has been observed in STM.
	Among the proposed explanations for a finite $dI/dV$ signal at the Fermi energy are defect states and substrate effects \cite{Crommie2017-SI}, a charge density wave \cite{Jia2017-SI} and a Coulomb gap \cite{Song2018-SI}, which we discuss below.
	By comparing the different STM studies of ML WTe$_2$ \cite{Jia2017-SI,Crommie2017-SI,Song2018-SI}, defect states seem to be an unlikely explanation because the non-zero $dI/dV$ signal is observed even far away from topographic defects and the reported defect concentrations vary from study to study. 
	While a CDW may explain the observed gap feature, the doping dependence of the band gap feature reported in Ref. \citenum{Song2018-SI} does not line up with a typical CDW behavior. Furthermore, the topographic CDW features are not well reproduced in reports other than in Ref. \citenum{Jia2017-SI}. 
	Lastly, a Coulomb gap due to lateral hopping in ML~WTe$_2$ was proposed as a possible explanation \cite{Song2018-SI}. This explanation is based on ML WTe$_2$ being decoupled from the substrate, which seems reasonable due to the van der Waals stacking and small Fermi surface overlap of ML~WTe$_2$ and graphene as well as NbSe$_2$. To understand the origin of a possible Coulomb gap in more detail, we consider tip-induced band bending (TIBB) to play a crucial role. 
	TIBB can have a significant effect on the measured $dI/dV$ spectra and is known to lead to a finite $dI/dV$ signal in the band gap \cite{Fennstra1994-SI}.
	The 2D analog of 3D TIBB is a local band shift in the 2D layer. As a result, there is no additional confinement of states in the $z$-direction but only in the $x$-$y$ plane, i.e. a laterally extended circular potential well.
	The depth of the potential well increases with decreasing tip-sample distance and decreasing STM tip work function and results in a local accumulation of charges at the position of the STM tip, i.e. a downward band shift.
	In the context of TIBB, the observed dip in the finite $dI/dV$ signal at the Fermi energy can be explained as a Coulomb gap associated with the accumulation layer that is formed in the potential well. 
	TIBB may also explain why the in-gap $dI/dV$ signal and the signature of the Coulomb gap is weak in Ref. \citenum{Crommie2017-SI}: in their experiments the current setpoint was lower (and the work function of their tip may have been higher), producing less TIBB and therefore fewer electrons in the accumulation layer compared to our results and Refs. \citenum{Jia2017-SI,Song2018-SI}.  Furthermore, the presence of a tip-induced electron accumulation layer in the WTe$_2$ can explain the reported conduction and valence band overlap extracted from quasi-particle interference (QPI) measurements in Ref. \citenum{Song2018-SI} which is in contrast to ARPES results \cite{Crommie2017-SI}: 
	with the finite $dI/dV$ signal in the band gap originating from the tip-induced electron accumulation layer, it is expected that the QPI of these states will yield wave vectors corresponding to the down-shifted conduction band states.
	As a result, QPI is expected to show conduction band wave vectors down to voltages beyond the actual conduction band edge and persist until they are dominated by the valence band conductance.
	In summary, we believe that TIBB may explain the majority of gap features reported in the literature.
	
	%%%%%%%%%%%%%%%%%%%%%%%%%%%%%%%%%%%%%%%%%%%%%%%%%%%%%%%%%%%%%%%%%%
	\begin{figure*}[hb!]
		\centering
		\includegraphics[width=0.9\textwidth]{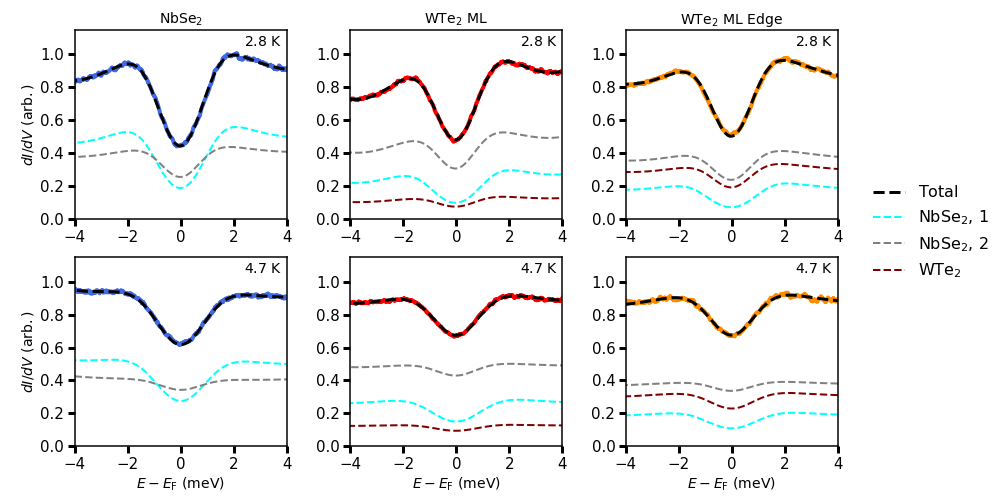}
		\caption{\label{fig:proximity_fits} {\bf Fits of the superconducting gaps.} Representative raw data differential conductance spectra of NbSe$_2$ (left), WTe$_2$ ML away from the edge (center), and at the WTe$_2$ ML edge (right) at STM baseplate temperatures of \SI{2.8}{K} (top) and \SI{4.7}{K} (bottom). Best fits of Eq. \ref{Eqn:WTe2-gap} are shown as the thick black dashed lines. Thinner dashed lines indicate the relative contribution of each term in Eq. \ref{Eqn:WTe2-gap}. All spectra taken at $V_\mathrm{sample}=\SI{5}{mV}$, $I_\mathrm{t}=\SI{100}{pA}$ and $V_\mathrm{mod} = \SI{100}{\murm V}$. In Fig.~\ref{Fig4}d and e of the main text, the two NbSe$_2$ contribution terms are combined for clarity. 
		}
	\end{figure*}
	%%%%%%%%%%%%%%%%%%%%%%%%%%%%%%%%%%%%%%%%%%%%%%%%%%%%%%%%%%%%%%%%%%

	\section{Spectra fitting}
	\label{spectra_fitting}
	The finite-temperature differential conductance function that we fit to the spectroscopic data is the density of states convolved with the derivative of the Fermi function, i.e.
	\begin{equation}
	\frac{dI}{dV}\propto\int_{-\infty}^{\infty}dE \frac{\exp\left(\frac{E-eV}{k_{\rm B}T_{\rm tip}}\right)}{\left[\exp\left(\frac{E-eV}{k_{\rm B}T_{\rm tip}}\right) + 1\right]^2}N(E)
	\label{Eqn:dIdV}
	\end{equation}
	where $N(E)$ is the density of states of the sample and the density of states in the tip is assumed to be uniform. In standard BCS theory, the density of states $N$ is given by
	\begin{equation}
	N(E, \Delta) = N_0{\rm Re}\left[\frac{E}{\sqrt{E^2 - \Delta^2}}\right]
	\label{Eqn:BCS}
	\end{equation}
	where $\Delta$ is the gap size of the superconductor and $N_0$ is the normal state density of states, which is typically assumed to be constant. However, in experiments the differential conductance signal often shows a non-constant normal state conductance around the Fermi energy. To achieve representative fits of the superconducting gaps, we therefore allow for a normal state conductance background given by a low-order polynomial function which we fit simultaneously with the BCS  gap function (this is then divided out of the superconducting gap spectra presented in the main text Figs. \ref{Fig2}, \ref{Fig3}, and \ref{Fig4}). We fit the NbSe$_2$ spectra using a two-gap model \cite{Khestanova2018-SI}, which we find provides good fits at all temperatures. The density of states for the two-gap model is given by
	\begin{equation}
	N_{\rm NbSe_2}(E, \Delta_1, \Delta_2)= CN(E, \Delta_1) + (1-C)N(E, \Delta_2)
	\label{Eqn:Two-gap}
	\end{equation}

	First, we fit data of Ref.~\citenum{Guillamon2008-SI} acquired at \SI{100}{mK}, yielding gaps of $\Delta_{\rm NbSe_2, 1} = \SI{1.1844 \pm 0.0021}{meV}$ and $\Delta_{\rm NbSe_2, 2} = 0.8300 \pm 0.0027$ meV (and a tip temperature of $T_{\rm tip} =1.0445 \pm 0.0134\rm\,K$). The amplitudes of the large-gap and small-gap terms are expressed by parameters $C$ and $(1-C)$, with a value of $C=0.5424 \pm 0.0047$ found from this fit. This same value of $C$ is then used in fits to our own NbSe$_2$ spectra, shown in Fig.~\ref{fig:proximity_fits}. The resulting gap sizes are $\Delta_{\rm NbSe_2, 1}=1.154 \pm 0.031$ meV and $\Delta_{\rm NbSe_2, 2} = 0.757 \pm 0.011$ meV with tip temperature of $T_{\rm tip}=5.962 \pm 0.033\rm\,K$ at a base plate temperature of \SI{2.8}{K}, and gaps of $\Delta_{\rm NbSe_2, 1} = 1.051 \pm 0.030$ and $\Delta_{\rm NbSe_2, 2} = 0.495 \pm 0.060$ meV with tip temperature of $T_{\rm tip}=7.405 \pm 0.052$ at a base plate temperature of \SI{4.7}{K}. We find that the temperature dependence of the larger of these NbSe$_2$ gap values matches very well to the universal temperature behavior of a BCS superconducting gap. The temperature dependence of the smaller gap, however, does not agree as well with the universal BCS prediction. The sample temperatures in our experiments, extracted from the known temperature dependence of the NbSe$_2$ gaps, are approximately \SI{1}{K} above our measured STM base plate temperatures.
	
	\begin{table*}[h]
		\centering
		\begin{tabular}{ |p{3.5cm}|P{3cm}|P{3cm} P{3cm}|  }
			\hline
			& Parameter & 2.8 K& 4.7 K\\
			\hline
			NbSe$_2$   & $\Delta_{\rm NbSe_2, 1} \rm\,(meV)$ & $1.154 \pm 0.031$   & $1.051\pm0.030$\\
			&   $\Delta_{\rm NbSe_2, 2}\rm\,(meV)$ & $0.757\pm0.011$ & $0.495\pm0.060$\\
			&$C$ & \multicolumn{2}{c|}{$0.5424 \pm 0.0047$ (fixed)}\\
			& $T_{\mathrm{tip}}\rm\,(K)$ & $5.962\pm0.033$ & $7.405\pm0.052$\\ \hline
			
			WTe$_2$ monolayer  & $\Dbulk\rm\,(meV)$   &  $0.83 \pm 0.08$   & $0.76 \pm 0.16$\\
			&$f_{\rm NbSe_2}$ & \multicolumn{2}{c|}{$0.86 \pm 0.04$}\\
			&$C$ & \multicolumn{2}{c|}{$0.33 \pm 0.06$}\\
			\hline
			
			WTe$_2$ ML edge & $\Dedge\rm\,(meV)$    &   $0.75 \pm 0.08$ &  $0.77 \pm 0.08$\\
			&$f_{\rm NbSe_2}$  & \multicolumn{2}{c|}{$0.65 \pm 0.06$}\\
			&$C$ &  \multicolumn{2}{c|}{$0.33 \pm 0.06$ (fixed)}\\
			\hline
			
			WTe$_2$ bilayer& $\DbulkBL\rm\,(meV)$    &    & $0.60 \pm 0.19$\\
			&$f_{\rm NbSe_2}$ &  \multicolumn{2}{c|}{$\rm{(min)} 0.25 - \rm{(max)}1.00$}\\
			&$C$ & \multicolumn{2}{c|}{$\rm{(min)}0.10 - \rm{(max)}0.33$}\\
			\hline
			WTe$_2$ three layers& $\Delta_{\rm{WTe_2}}^{\rm(3L)}\rm\,(meV)$    &    & $0.37\pm 0.05$\\
			\hline
			
			WTe$_2$ five layers& $\Delta_{\rm{WTe_2}}^{\rm(5L)}\rm\,(meV)$     &    & $ 0.31\pm0.03$\\
			\hline
			
			WTe$_2$ six layers& $\Delta_{\rm{WTe_2}}^{\rm(6L)}\rm\,(meV)$    &    & $ 0.31\pm0.02$\\
			\hline
			
			WTe$_2$ seven layers & $\Delta_{\rm{WTe_2}}^{\rm(7L)}\rm\,(meV)$    &    & $ 0.36\pm0.03$\\
			\hline
		\end{tabular}
		\caption{\textbf{Summary of fitting results.} Details described in the text.
		}
		\label{tab:fit_results}
	\end{table*}
	
	\vspace{3mm}
	
	\textbf{Fitting of WTe$_2$ monolayer and monolayer edge on NbSe$_2$.} Moving to the spectra acquired on the monolayer and the monolayer edge of WTe$_2$ on NbSe$_2$ and comparing those spectra to ones acquired on WTe$_2$ on hBN (Fig.~\ref{Fig4} of the main text), we find that there is a significant contribution to the tunneling current from the underlying NbSe$_2$. We therefore fit our ML WTe$_2$ spectra with a three-gap model, two gaps of known size from the NbSe$_2$ and one induced gap from the WTe$_2$, $\Dbulk$, the size of which is determined by fitting
	\begin{equation}
	N(E)= (1-f_{\rm NbSe_2})N(E, \Dbulk) + f_{\rm NbSe_2}N_{\rm NbSe_2}(E, \Delta_{\rm NbSe_2, 1}, \Delta_{\rm NbSe_2, 2}).
	\label{Eqn:WTe2-gap}
	\end{equation}
	We fix the fractional contribution of the NbSe$_2$, $f_{\rm{NbSe}_2}\equiv B/(A+B)$, by comparing the value of the observed conductance at the Fermi energy for ML WTe$_2$ on NbSe$_2$ and for ML WTe$_2$ on hBN. Importantly, for the NbSe$_2$ contribution, the relative amplitudes of the large-gap and small-gap terms is expected to differ from the $C=0.5424 \pm 0.0047$ deduced above from bare NbSe$_2$, since the wave functions of the respective states will have varying decay constants as they extend through the WTe$_2$ and out into the vacuum. We therefore let $C$ vary in the fits of the WTe$_2$ spectra. The tip temperatures are known from the fits to the bare NbSe$_2$ spectra and are fixed for fits of the spectra taken on the WTe$_2$.

	To maximally constrain the fits, we simultaneously fit the spectra acquired at temperatures \SI{2.8}{K} and \SI{4.7}{K}, thus having three unknowns in total to fit the two spectra: the induced superconducting gap size for each, and the relative amplitude of the NbSe$_2$ large-gap and small-gap terms (i.e. the same value in both spectra). Using the best fit, midpoint values for all other ``auxiliary'' parameters (NbSe$_2$ gap sizes, sample temperatures, WTe$_2$ fractional contribution), we find values for the proximity-induced gaps in the WTe$_2$ of $\SI{0.80 \pm 0.04}{meV}$ (at \SI{2.8}{K}) and $\SI{0.73 \pm 0.04}{meV}$ (at \SI{4.7}{K}) and $C=0.35 \pm 0.02$ (Fig.~\ref{fig:proximity_fits}). To properly estimate error bounds for the induced gap, we further perform fits at the minimum and maximum values of the error range of all auxiliary parameters, thus yielding final values for the proximity-induced gaps for ML WTe$_2$ far from an edge of $\Dbulk = \SI{0.83 \pm 0.08}{meV}$ and $\Dbulk = \SI{0.76 \pm 0.16}{meV}$, respectively at $\SI{2.8}{K}$ and $\SI{4.7}{K}$, and $C=0.33 \pm 0.04$. 
	For the case of the WTe$_2$ ML edge on NbSe$_2$, we fit a spectrum acquired at \SI{2.8}{K} using the known NbSe$_2$ gaps and known tip temperature, with $f_{\rm NbSe_2}$ determined to be $0.65 \pm 0.06$ by comparison of the edge spectrum (Fig.~\ref{Fig2}b) with those of the non-edge WTe$_2$ on NbSe$_2$ and WTe$_2$ on hBN (we make this indirect comparison because it is not possible to measure ML edge spectra on hBN, which would require there to be an exposed insulating hBN surface). We fit the edge spectrum using the value of $C=0.33 \pm 0.04$, which we obtained from the monolayer fits. The resulting fit is shown in Fig.~\ref{fig:proximity_fits}, with induced gap of $\Dedge = \SI{0.75 \pm 0.08}{meV}$. For the case of acquisition temperature \SI{4.7}{K} (not shown), the same procedure yields an induced gap of $\Dedge = \SI{0.77 \pm 0.08}{meV}$. 
	We note that we are not able to achieve good fits of our WTe$_2$ data if we assume a constant WTe$_2$ contribution, i.e. one that does not have a superconducting gap.
	\\
	\begin{figure*}[ht!]
		\centering
		\includegraphics[width=12cm]{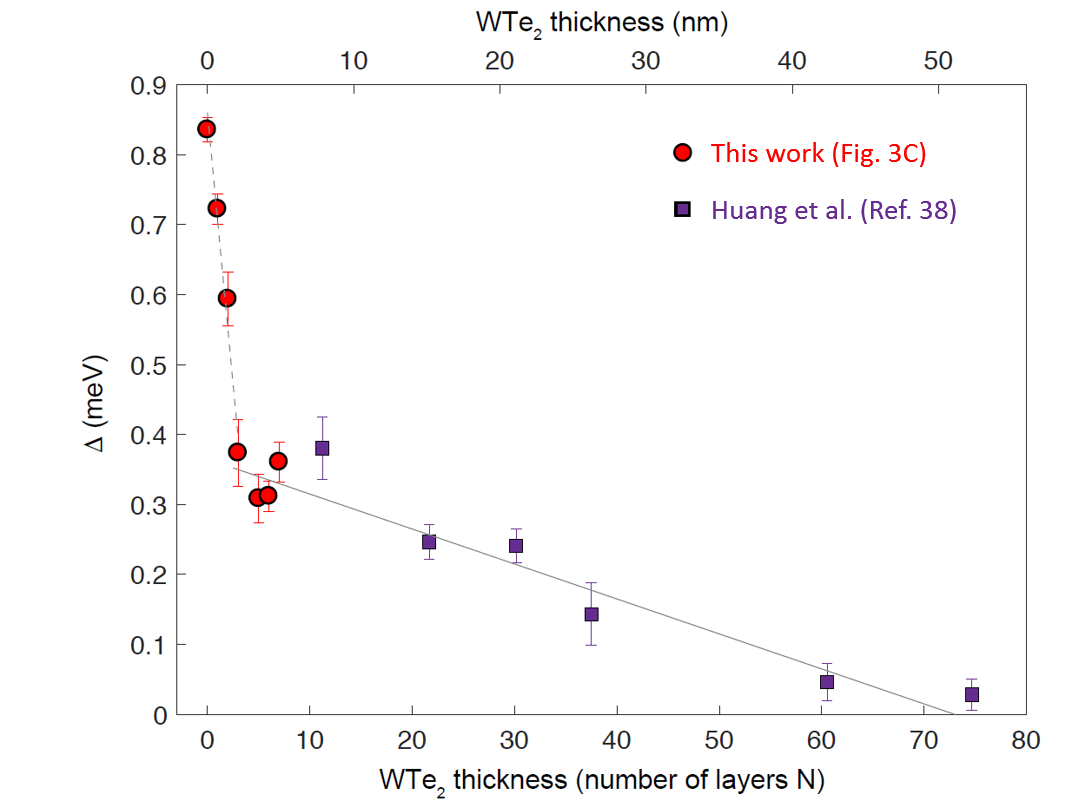}
		\caption{\label{Fig:Delta_v_thickness}
			{\bf Comparison of observed superconducting gap thickness dependence to the literature.} Observed superconducting gap $\Delta$ vs. layer number $N$, from Fig.~\ref{Fig3}c of the main text (red circles) plotted alongside data extracted from transport measurements in Ref.~\citenum{Huang2018-SI} (purple squares). The grey dashed and grey solid lines, which are guides to the eye, show the trend of the observed gap size in the `thin' limit ($N<3$ layers) and `thick' ($N\geq3$ layers) limit respectively. The purple data points are calculated by extracting the $\Delta/\Delta_{\rm{NbSe_2}}$ values from Fig.~5 in Ref. \citenum{Huang2018-SI}, taken at $\approx\SI{2}{K}$, and multiplying them by $\Delta_{\rm{NbSe_2}}\approx 0.6\,\rm{meV}$ at \SI{4.7}{K}, a number obtained from Fig.~2g of the same paper ($\Delta_{\rm{NbSe_2}} \equiv \Delta_0$ in that paper).  This procedure is valid because the ratio $\Delta/\Delta_{\rm{NbSe_2}}$ has a very weak temperature dependence (Fig.~2g of Ref.~\citenum{Huang2018-SI}). This allows for direct comparison with our data, which were obtained at \SI{4.7}{K}.
		}
	\end{figure*}

	\textbf{Fitting of multilayer WTe$_2$ on NbSe$_2$}. We now consider thicker WTe$_2$ on NbSe$_2$, for which we have measured tunneling spectra at \SI{4.7}{K} (Fig.~\ref{Fig3}). 
	For $N=3$ or more layers of WTe$_2$ we fit the observed spectra with a single-gap model (Eq. \ref{Eqn:BCS}), obtaining proximity-induced gaps in good agreement with the thickness-trend found in prior measurements of $N>10$ layers of WTe$_2$ on SiO$_2$ \cite{Huang2018-SI} as shown in Fig.~\ref{Fig:Delta_v_thickness}. We conclude that the tunneling contribution of NbSe$_2$ to the observed spectra is negligible for $N\geq 3$. For the case of bilayer WTe$_2$ on NbSe$_2$, as for monolayer WTe$_2$/NbSe$_2$, we must estimate the contribution of the NbSe$_2$ to the bilayer WTe$_2$/NbSe$_2$ spectrum. 
	To do so, we have measured a spectrum of bilayer WTe$_2$ on hBN (Fig.~\ref{Fig:WTe2-hBN}c), from which we find the tunneling conductance at the Fermi energy to be $1.8\times$ larger than that for ML WTe$_2$ on hBN, thus implying a \textit{lower bound} for the contribution of WTe$_2$ itself to the spectrum of bilayer WTe$_2$ on NbSe$_2$ of $1.8\times0.14=0.25$, with the additional conservative assumption that the \textit{total} tunneling current into the NbSe$_2$ on BL~WTe$_2$/NbSe$_2$ is the same as on ML WTe$_2$/NbSe$_2$ (in general we expect the total tunneling into BL~WTe$_2$/NbSe$_2$ to be less.)  We therefore consider a range of WTe$_2$ fractional contributions for the bilayer case extending from 0.25 to 1.0.
	For the fractional magnitude of the NbSe$_2$ large-gap term, in the discussion above we found values of $C = 0.54$ and $C=0.33$ for zero and one WTe$_2$ layer, respectively; thus, for two layers of WTe$_2$ we assume values in the range $C=0.10-0.33$. Within these parameter ranges, we find a proximity-induced gap for the bilayer WTe$_2$ to be in the range $\DbulkBL = {0.41 - 0.79}\rm\,{meV}$, or $\DbulkBL = \SI{0.60 \pm 0.19}{meV}$. Our fit results are summarized in Table~\ref{tab:fit_results}.

	\section{Density Functional Theory Calculations}
	\label{DFT}
	\begin{figure*}[h]
		\centering
		\includegraphics[width=0.6\textwidth]{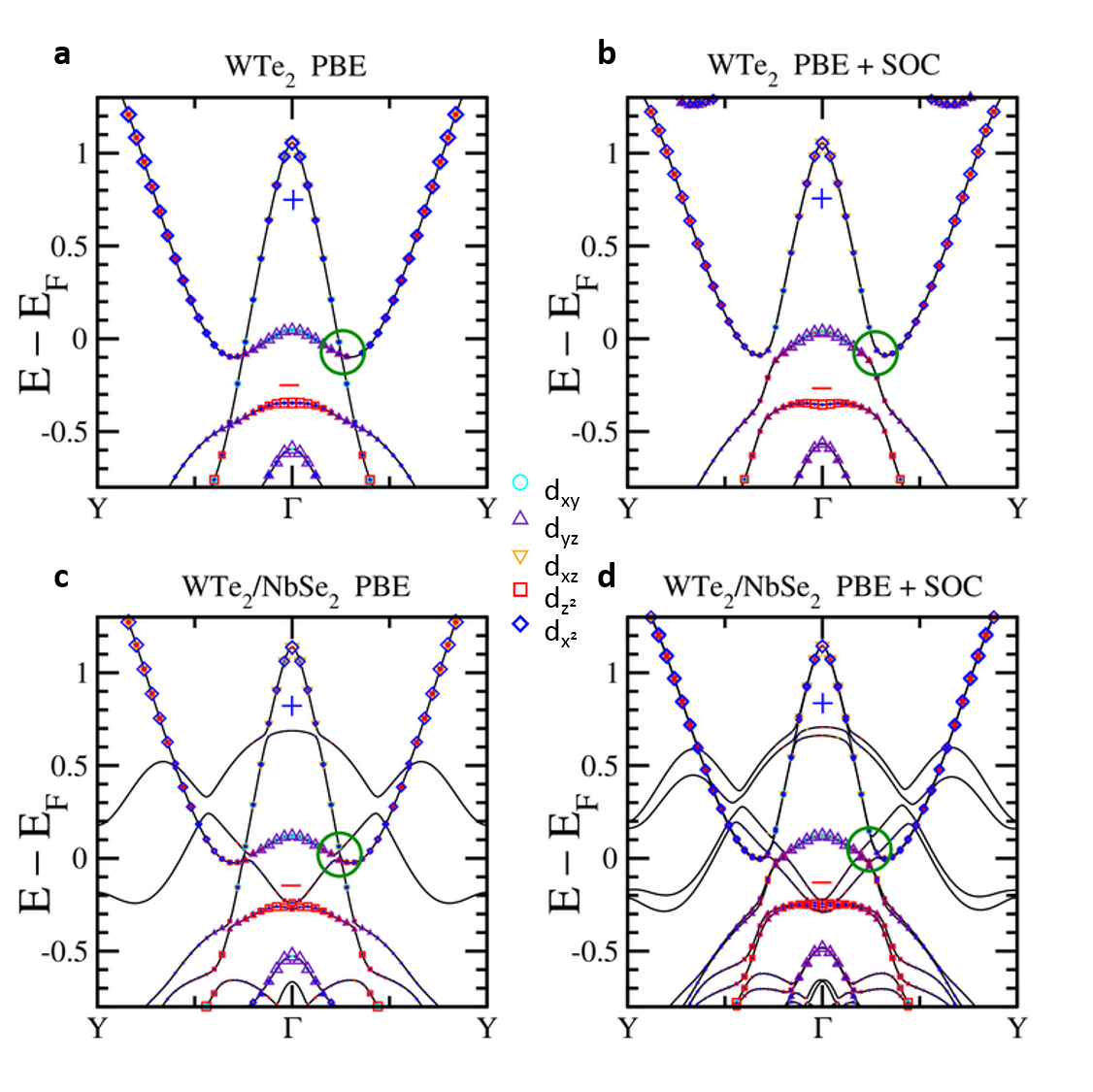}
		\caption{\label{FigS6} \textbf{DFT band structure calculations}. \textbf{(a)} ML 1T'-WTe$_2$ without spin-orbit coupling (SOC) and \textbf{(b)} with SOC. 
			\textbf{(c)} and \textbf{(d)} show the band structure for a heterostructure of ML 1T'-WTe$_2$ and a single layer of 2H-NbSe$_2$, respectively. Calculation details are described in the text. The green circle indicates where SOC forces bands to anti-cross in the pockets along the Y-$\rm{\Gamma}$-Y path for both the ML WTe$_2$ and the heterostructure. The blue $+$ and red $-$ indicates the parity of the states associated with the band inversion resulting in the QSH edge state, in comparison to Fig.~2d and e from Ref. \citenum{Crommie2017-SI}. 
		}
	\end{figure*}
	We performed band structure calculations (Fig.~\ref{FigS6}) using density functional theory (DFT) in the generalized gradient approximation \cite{Perdew1996-SI} utilizing the projector augmented wave method as implemented in VASP \cite{Kresse1999-SI}. Lattice constants  were set to orthorhombic WTe$_2$ values ($a=3.477\,$\AA\ and $b=6.249\,$\AA) for both the ML and the WTe$_2$/NbSe$_2$ bilayer, while we took $c=30\,$\AA\ to allow sufficient vacuum. Atomic coordinates were relaxed holding the lattice constants fixed prior to calculating the band structure.
	The results for the freestanding WTe$_2$ monolayer, with and without spin-orbit-coupling (SOC) are shown in Fig.~\ref{FigS6}a and b. Note that in WTe$_2$ the bands near the Fermi energy originate mostly from the W atoms, the orbital character of which we plot as symbols as indicated in the figure legend.
	The calculations are in excellent agreement with previously reported band structures of 1T'-WTe$_2$. Especially, the inversion of bands with opposite parity is reproduced (marked as "+" and "-" in the figures). Note that the size of the gap, which opens when including SOC into the calculations is under predicted in DFT.
	
	Next, we perform calculations which, in addition to the ML WTe$_2$, include a layer of NbSe$_2$ to model the present sample system. For this purpose, we construct a heterostructure unit cell which includes $2\times1$ unit cells of NbSe$_2$ and $1\times1$ unit cells of WTe$_2$. The NbSe$_2$ lattice parameter is adjusted such that the constructed unit cells match in size.
	We then let the structure relax, before calculating the supercell band structure, with and without SOC.
	Figures~\ref{FigS6}c and d show the resulting heterostructure bands, where we plot the W orbital character as symbols as before. The NbSe$_2$ bands near the Fermi energy are dominated by Nb which we plot without symbols.
	From the calculations, we find that the NbSe$_2$ has little effect on the the WTe$_2$ single-particle electronic structure, i.e. we observe no significant shifts of bands relative to each other and the band inversion is still present. 
	There is only a small rigid upward shift of the WTe$_2$ bands ($\SI{<100}{meV}$) with respect to the calculated freestanding WTe$_2$ due to an NbSe$_2$ band near the Fermi level. Furthermore, we observe a slight hybridization between bands that originate in the NbSe$_2$ and the even parity conduction band of the WTe$_2$, but only at an energy well above the Fermi level. Note that the band splitting in Fig.~\ref{FigS6}d comes from the broken inversion symmetry due to the presence of the NbSe$_2$ which lifts the spin degeneracy of the bands when SOC is included.
	In conclusion, we find that the WTe$_2$ single-particle band structure, including the band inversion and therefore the QSH state, is preserved in the WTe$_2$/NbSe$_2$ heterostructure. 
	%\clearpage

\end{document}